\documentclass[aps,pre,twocolumn]{revtex4-1}
\usepackage{amssymb}
\usepackage{amsmath,bm}
\usepackage{graphicx,color}
\usepackage{bbm}
\usepackage{multirow}
\newcommand{\B}[1]{{\bm{#1}}}

\begin{document}

\title{Plastic Instabilities in Charged Granular Systems: Competition between Elasticity and Electrostatics}
\author{Prasenjit Das$^1$, H. George E. Hentschel$^{1,2}$ and Itamar Procaccia$^{1,3}$}
\affiliation{$^1$Department of Chemical Physics, The Weizmann Institute of Science, Rehovot 76100, Israel.\\
$^2$ Dept. of Physics, Emory University, Atlanta Ga. 30322, USA.\\ $^3$  Center for OPTical IMagery Analysis and Learning, Northwestern Polytechnical University, Xi'an, 710072 China.  }

\begin{abstract}
Electrostatic theory preserves charges, but allows dipolar excitations. Elasticity theory preserves dipoles, but allows quadrupolar (Eshelby like) plastic events. Charged amorphous granular systems are interesting in their own right; here we focus on their plastic instabilities and examine their mechanical response to external strain and to external electric field, to expose the competition between elasticity and electrostatics. In this paper a generic model is offered, its mechanical instabilities are examined and a theoretical analysis is presented. Plastic instabilities are discussed as saddle-node bifurcations that can be fully understood in terms of eigenvalues and eigenfunctions of the relevant Hessian matrix. This system exhibits moduli that describe how electric polarization and stress are influenced by strain and electric field. Theoretical expression for these moduli are offered and compared to the measurements in numerical simulations. 
\end{abstract}

\maketitle

\section{Introduction}
Charged granular amorphous matter abounds in nature and technology, from wind-blown sand \cite{08KR} and colliding granular matter \cite{89NIS,18SM}, to dense powders \cite{05LH,18JSTW} and astrophysical dust \cite{00Pop}. The physics of such states of matter are interesting and rich, having attracted considerable amount of research, Cf. \cite{99WSS,00Igor}. Our focus in this paper is on their mechanical properties under external mechanical strains and external electric fields, and on their plastic responses. These are particularly interesting since charged granules exert standard (binary) elastic short range forces upon contact, simultaneous with long range electric forces. The plastic responses of amorphous solids with short range forces have been studied extensively in recent years; having elastic properties at small external strains, their plastic responses are dictated by elasticity theory \cite{18Lem} which preserves dipoles. This property stems from the bi-Laplacian nature of elasticity theory, in contrast to electrostatics that is a Laplacian theory that preserves charges. Therefore quite generically the plastic responses in which stress and mechanical energy are lost are associated with quadrupolar displacement fields, known as ``Eshelby" responses after Ref.~\cite{57Esh}. As said, electrostatic theory on the other hand preserves charges, and therefore the ``cheapest" expected responses are dipoles rather than quadrupoles. It is thus timely and interesting to examine the fundamental plastic responses of compressed charged amorphous solids with the aim of discovering which type of interaction dominates their plastic events. In this paper we find that the quasi-localized plastic responses to strain and electric field are fully characterized by the eigenvalues and eigenfunctions of the appropriate Hessian matrix which is derived below.  We also discuss the generalized moduli that this kind of systems exhibit - especially the mixed moduli relating stress to electric field and electric polarization to strain and electric field, known as piezoelectricity and electrostriction. 

In Sect.~\ref{model} we describe the model studied below. The short-range repulsive forces and the long range electric forces are described. We stress that the model presented is simple, we do not consider surface charges which are induced by friction, and our granules are not deformable. In this sense the studies presented below are preliminary and much more work to model and understand charged sand, charged powders or charged colloids is called for. In Sect.~\ref{protocols} the straining protocols are introduced. These include mechanical shear straining on the one hand and increased electric field on the other. Both protocols lead to elastic responses interspersed with sharp plastic failures. The theory required to understand the nature of the irreversible non-affine response of the system is developed in Sect.~\ref{theory}. Sect.~\ref{moduli} deals with the elastic responses, both due to increased strain and growing electric field. This system requires a host of moduli which related the increase of stress due to mechanical strain and increased electric field, but also the reversible response of electric polarization to the same mechanical strain and electric field. In Sect.~\ref{summary} we present a short summary and some conclusions.

\section{The model system}
\label{model}
\begin{figure}
    \includegraphics [scale=0.53]{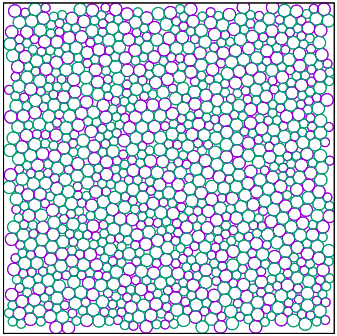}
    \caption{Typical initial configuration of a system which is later subjected to a simple shear. Grains with positive and negative charges are, respectively, marked in blue and green. Sizes of circles are proportional to the disks diameters.}
     \label{config}
\end{figure}

In powders and sand charges accumulate on the surface of granules. Here we construct a simpler model with charges fixed on the centers of mass, eliminating charge transfer from one granule to another. To examine the plastic events in charged compressed granular media we study a model consisting of a 50-50 mixture of frictionless 2-dimensional disks with diameters $R_1=1.0$ and $R_2=1.4$ respectively. Half of the small particles are positively charged at the center of mass with a charge $+q$ and the other half are negatively charged with a charge $-q$. The same is true for the large particles. The disks are placed randomly inside a two-dimensional box of size $L_{in}=50R_1$ such that there is no overlap between two particles and the system is equilibrated using molecular dynamics with global damping, meaning that to each equation of motion one adds a damping term $-\kappa \dot {\B r_i}$ . Then we compress the system quasistatically to achieve a required packing fraction $\phi>\phi_J$  where $\phi_J \approx 0.843$ is the jamming packing fraction at zero temperature (for uncharged systems). This way we form an amorphous solid that is charge-neutral. A typical initial configuration of a system is shown in Fig.~\ref{config}. In this example the compressed box length is $L=\sqrt{A/\phi}$, where $A$ is the total area covered by the disks. All the lengths are measured in the unit of $R_1$. The simulation presented below employs periodic boundary conditions, the total number of particles is $N=1000$, $\phi=0.90$, and therefore $L=35.938053R_1$.

The short-ranged forces between two overlapping disks are Hertzian-elastic. The potential for these forces is given by \cite{01SEGHLP}:
\begin{equation}
	\label{eqn1}
	\varPhi_{elas}(r_{ij}) = \frac{2}{5}K_n\sqrt{R_{eff}}(R_{ij}-r_{ij})^{5/2}.
\end{equation}
Here, $K_n$ is an elastic constant. Denoting the centers of mass of the $i$th and $j$th disk as $\B r_i$ and $\B r_j$ then $r_{ij} = |\B r_i-\B r_j|$, $R_{ij} =  (R_i + R_j)/2$ and $R_{eff} = 0.5R_i R_j/(R_i+R_j)$. We do not consider any kind of damping to the elastic force since we strain the systems quasi-statically at zero temperature.

Apart from the elastic force, grains interact via long-ranged electrostatic forces. If $q_i$ and $q_j$ are the charges in the $i^{th}$ and $j^{th}$ grains, the electrostatic interaction potential is given, in Gaussian units,  by
\begin{align}
\label{eqn2}
\tilde V_{\rm elec}(r_{ij}) = \frac{q_iq_j}{ r_{ij}},
\end{align}
In our simulation we use units of charge such that $q_i=\pm 1$. The electrostatic interaction is of course long-ranged. However, it has been shown \cite{06FG,07CBHIK} that in an amorphous mixture of randomly distributed charged grains, one can use the damped-truncated Coulomb potential as given by
\begin{align}
\label{eqn3}
V_{\rm elec}(r_{ij}) =  q_iq_j\left[ \frac{erfc(\alpha r_{ij})}{r_{ij}} - \frac{erfc(\alpha R_c)}{R_c} \right], r_{ij}\leq R_c.
\end{align}
Here $erfc(x)$ is the complementary error function, $\alpha$ is the damping factor of the electrostatic interaction due to screening and $R_c$ is the cut-off range of the electrostatic interaction. Below we employ the Hessian matrix, which is the second derivative of the potential with respect to coordinates. We therefore smooth out $V_{elec}$ at $r=R_c$ to have four derivatives when $V_{\rm elec}$ goes to zero at $r=R_c$. To this aim we use the following form
\begin{align}
	\label{eqn4}
	\varPhi_{\rm elec}(r_{ij}) = V_{\rm elec}(r_{ij}) - \sum_{n=1}^{4}\frac{(r_{ij} - R_c)^n}{n!}\left.\frac{d^nV_{\rm elec}}{dr_{ij}^n}\right|_{r_{ij}=R_c}.
\end{align}
It should be stressed at this point that we have tested whether using the full interaction range may introduce any change in the results presented below. The answer is no, except forcing much longer simulations. We did not observe any qualitative change and only very minor quantitative changes.

Below we use the total binary potential $\varPhi(r_{ij})$ according to:
\begin{equation}
\varPhi(r_{ij}) \equiv	\varPhi_{\rm elas}(r_{ij})+ \varPhi_{\rm elec}(r_{ij}) \ .
\label{defphi}
\end{equation}

The total energy of the system in the presence of an electric field $\B E$ and strain $\B \gamma$ is then
\begin{equation}
U(\{\B r_i(\B \gamma,\B E)\}_{i=i}^N; \B E) = \sum_{i\le j} \varPhi(r_{ij}(\B \gamma,\B E))  - \sum_i q_i\B r_i(\B \gamma,\B E)\cdot \B E \ . 
\label{defU}
\end{equation}

In the numerical simulation which are reported next we set $K_n=20000$, $\alpha=0.1$ and $R_c=12.5$. The state of the system will be monitored below by measuring the stress $\B \sigma$ and the polarization $\B P$. These are defined as follows
\begin{equation}
\sigma^{\alpha\beta} \equiv \frac{1}{2L^2} \sum_{i,j=1}^N f_{ij}^\alpha r_{ij}^\beta\ , \quad
P^\alpha \equiv \frac{1}{L^2}\sum_{i=1}^N q_i r_i^\alpha \ .
\end{equation}
Here the force $\B f_{ij}$ exerted by particle $j$ on particle $i$ is
\begin{equation}
f^\alpha_{ij} = -\frac{\partial \varPhi(r_{ij})}{\partial r^\alpha_{ij}} \ .
\end{equation}

\section{Straining Protocols and Plastic Responses}
\label{protocols}
\begin{figure}
	\vskip 0.5 cm
	\includegraphics[width=0.70\columnwidth]{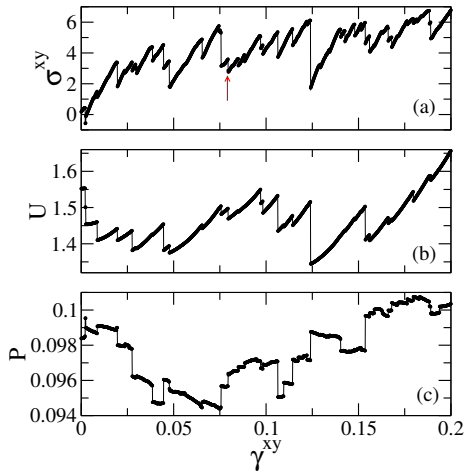}
	\caption{Panel (a): A typical graph of stress vs. strain as obtained with an AQS straining protocol without electric field. Note that the smooth elastic segments end with a plastic discontinuity. Panel 	(b): Total energy (elastic and electrostatic) as a function of strain. Panel (c): Polarization as a function of strain. Note that whereas all the plastic discontinuities in strain and total energy are negative drops, the polarization has both negative and positive jumps. In this simulation $P^x$ is about twice the size of $P^y$.}
    \label{Uvsgam}
\end{figure}

\begin{figure}
	\includegraphics[width=0.70\columnwidth]{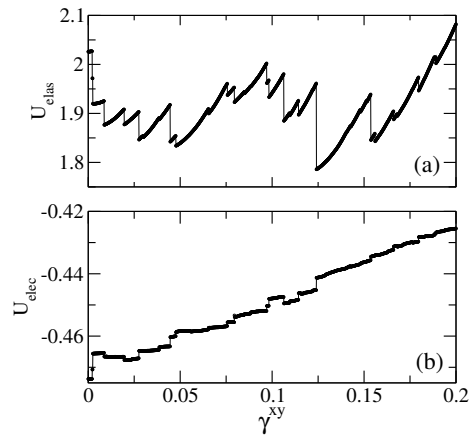}
	\caption{Panel (a): The elastic energy contribution to panel (b) in Fig.~\ref{Uvsgam}. Panel (b): The electrostatic energy contribution to panel (b) in Fig.~\ref{Uvsgam}. Since at the present parameters the elastic energy drops are bigger than the gains exhibited by the electrostatic energy, the total energy exhibits only drops at the plastic events. }
	\label{esvsgam}
\end{figure}

To examine the plastic responses of the model to {\em mechanical} strains, we impose on the system simple shear-strain in a quasi-static manner.  Here ``quasi-static" means that after every small step of strain we equilibrate the system by gradient energy minimization. The affine step of straining is achieved by increasing in  the strain by $\Delta\gamma^{xy}$. Here the electric field is zero. The straining step is defined by the volume preserving transformation
\begin{eqnarray}
x^{\prime}_i&=&x_i+\Delta\gamma^{xy} ~ y_i\nonumber \\
y^{\prime}_i&=&y_i.
\label{simpleshear}
\end{eqnarray}
This strain step is known as the ``affine" part in the system displacement. Due to the gradient energy minimization the system will also have a ``non-affine" displacement field that we denote as $\B u$.  We apply Lees-Edwards boundary conditions to the particle positions. We calculate the stress $\sigma^{xy}$, the total energy $U$ and the magnitude of the polarization $ P=|\B P|$ after each equilibration step as a function of $\gamma^{xy}$. Typical results are shown in Fig.~\ref{Uvsgam}. Here the size of the strain step is $\Delta\gamma^{xy}=10^{-4}$.
We note that smooth elastic increases in stress and  energy during the straining are interspersed with sharp non-affine plastic drops. These are of course irreversible. In contradistinction, the polarization experience both positive and negative jumps at the plastic events. This different character is underlined by splitting the energy into its elastic and electrostatic contributions, cf. Fig~\ref{esvsgam}. Here the elastic part of the total energy is $U_{\rm elas}\equiv \sum_{i\le j} \varPhi_{\rm elas}(r_{ij})$ and the electric part of the total energy $U_{\rm elec}\equiv \sum_{i\le j} \varPhi_{\rm elec}(r_{ij})$. While the elastic energy exhibits only losses in the plastic events (panel (a)), the electrostatic contribution to the energy can gain or lose upon a plastic event (panel(b)). 
Note that with the present parameters the elastic energy drops are bigger than the gains experienced by the electrostatic energy. Accordingly,  the total energy exhibits only drops at the plastic events.

\begin{figure}
	\includegraphics[width=0.70\columnwidth]{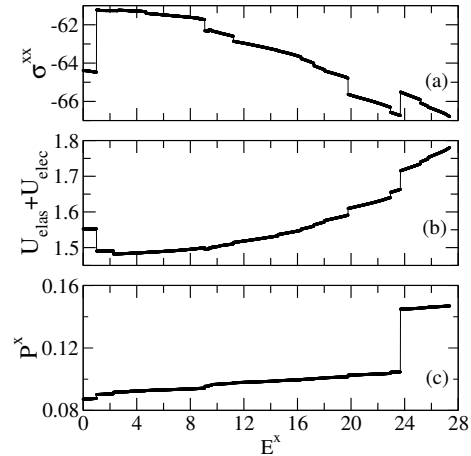}
	\caption{The stress, energy and polarization as a function of electric field. Note the sharp discontinuities which represent plastic events achieved by a saddle-node bifurcation. }
	\label{E}
\end{figure}

Upon increasing an external electric field in the absence of strain, the nature of the non-affine responses change. We increase the electric field in steps of $\Delta E^x=10^{-2}$ and minimize the energy using a FIRE algorithm \cite{06BKGMG}. Firstly, we learn that by increasing $E^x$ the change in $\sigma^{xy}$ is negligible, but $\sigma^{xx}$ responds strongly.  In Fig.~\ref{E} we present the electrostatic stress component $\sigma^{xx}$, $U_{\rm elas}+U_{\rm elec}$, and $P^x$ as a function of increasing electric field pointing in the $x$ direction. One observes again sharp changes at given values of the electric field, and as said, these necessitate a different theory from the plastic events under strain. 

\begin{figure}
	\includegraphics[width=0.70\columnwidth]{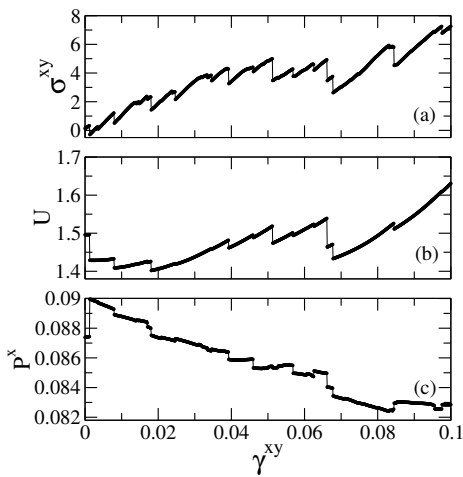}
	\caption{The stress $\sigma^{xy}$, total energy $U$ and polarization $P^x$ as a function of strain in the presence of electric field. The sharp discontinuities are again achieved by a saddle-node bifurcation. The present value of electric field is $E^x=0.5$.}
	\label{withE}
\end{figure}

Finally, we examine the responses to increasing strain with the presence of electric field. We first increase the electric field to $E^x=0.5$ in steps of $\Delta E^x=10^{-2}$ and then apply the strain in steps of $\Delta \gamma^{xy} = 10^{-4}$. We use again FIRE algorithm to minimize the energy after every step. Representative results are shown in Fig.~\ref{withE}.

\section{Theory} 
\label{theory}

To understand the nature of the plastic events one employs the Hessian matrix which is the second derivative of the energy function Eq.~(\ref{defU}) with respect to coordinates. One notes that the last term in Eq.~(\ref{defU}) does not contribute to this second derivative and we therefore can employ the part of the energy that depends on coordinate differences only and write for $i\ne j$ \cite{LemaitreMaloney06}:
\begin{equation}
H_{ij}^{\alpha\beta}\!=\!-\Big(\frac{\partial^2\Phi(r_{ij})}{\partial r_{ij}^2} \! -\! \frac{1}{r_{ij}}\frac{\partial\Phi(r_{ij})}{\partial r_{ij}} \Big)n_{ij}^\alpha n_{ij}^\beta
- \frac{\delta_{\alpha\beta}}{r_{ij}}\frac{\partial\Phi(r_{ij})}{\partial r_{ij}} \ ,
\end{equation}
where $n_{ij}^\alpha=(r_j^\alpha - r_i^\alpha)/r_{ij}$. The diagonal elements of the Hessian matrix read
\begin{equation}
H_{ii}^{\alpha \beta} =- \sum_{\ell\ne i} H_{i\ell}^{\alpha \beta} \ .
\end{equation}

\begin{figure}
	\vskip 0.5 cm
	\includegraphics[width=0.70\columnwidth]{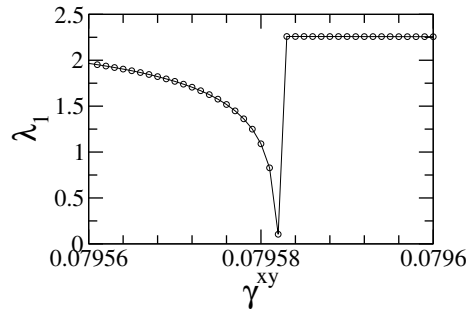}
	\caption{Plot of lowest non-zero eigenvalue $\lambda_1$ of the Hessian matrix as a function $\gamma^{xy}$ close to a plastic drop at $\gamma_p\approx 0.0795$ (red arrow in Fig.~\ref{Uvsgam} (a) ). In the straining protocol $\Delta\gamma^{xy}=10^{-6}$.}
	\label{eig1tozero}
\end{figure}

The Hessian matrix, being real and symmetric, has real eigenvalues. Besides Goldstone modes associated with continuous translational symmetries that yield two zero eigenvalues, all the other eigenvalues are positive as long as the system is mechanically stable. The instabilities are signaled by a positive eigenvalue going to zero at some strain value $\gamma_P$. As usual the eigenvalue approaches zero like $\sqrt{\gamma_P-\gamma^{xy}}$ due to the generic saddle node bifurcation that is associated with changing minima through crossing a saddle in the energy landscape \cite{99ML,MaloneyLemaitre06,12KLP}. An example of this is shown in Fig.~\ref{eig1tozero}. Instabilities under the increase of electric field are also saddle nodes as we show below. 

\begin{figure}
	\includegraphics[width=0.70\columnwidth]{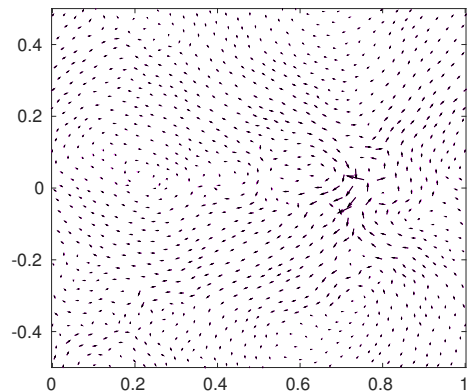}
	\caption{Plots of the non-affine displacement field $\B u$ (magenta) and the eigenfunction $\B \Psi_1$ associated with the lowest eigenvalue (black) for a characteristic 	plastic events. Here $\gamma_P=0.0795$ (Red arrow in Fig.~\ref{Uvsgam}); The dot product $|\B u \cdot\B \Psi_1|=0.994$. Note that the event appears quadrupolar as expected.}
	\label{compare}
\end{figure}

Sufficiently close to the saddle node bifurcation one expects the eigenfunction $\B \Psi_1$ associated with the lowest non-zero eigenvalue $\lambda_1$ to be very close to the non-affine displacement field $\B u$. The non-affine field is obtained by examining the last step of strain increase before the plastic discontinuity, and monitoring the displacement field resulting from the gradient energy minimization after the strain step. Subtracting from this displacement field the affine contribution Eq.~(\ref{simpleshear}) results in the non-affine displacement field $\B u$. In Fig.~\ref{compare} we show the eigenfunction superimposed on the normalized non-affine displacement. The closeness of the two fields is measured by the dot-product of the non-affine displacement and the eigenfunction $|\B u \cdot \B \Psi_1|$ that in the present case is 0.994. Note that the non-affine displacement field appears quadrupolar. This is a clear indication that the elastic interactions are dominant.

\begin{figure}
	\includegraphics[width=0.70\columnwidth]{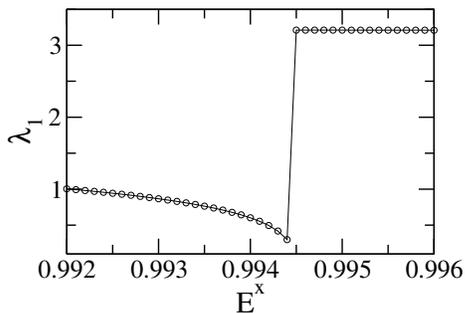}
	\caption{A typical change of the lowest non-zero eigenvalue associated with one of the sharp events shown in Fig.~\ref{E}. Here the steps of increasing in electric field are $\Delta E^x=10^{-4}$. }
	\label{lamE}
\end{figure}

\begin{figure}
	\includegraphics[width=0.75\columnwidth]{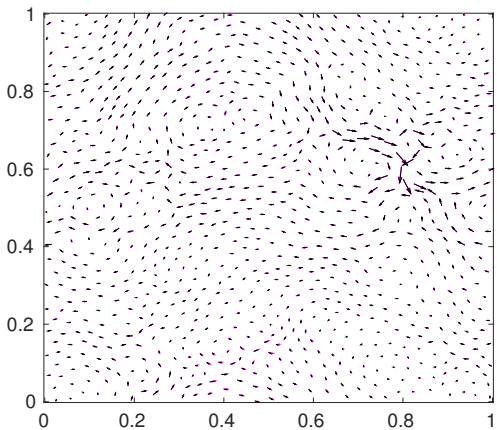}
	\caption{A typical non-affine response associated with the sharp jumps shown in Fig.~\ref{E} as a response to the increase in the electric field. The non-affine response is shown together with the eigenfunction associated with the lowest non-zero eigenvalue, cf. Fig.~\ref{lamE}. The dot-product of the non-affine displacement and the eigenfunction $|\B u \cdot \B \Psi_1|$ in the present case is 0.998. While the quadrupolar nature of the event is less pronounced than in the case of pure mechanical strain, it is still not a dipolar event.}
	\label{nonaffE}
\end{figure}

In this work we learn that the plastic events shown in Fig.~\ref{E} due to the electric field can be explained in the same manner. Again, the lowest eigenvalue of the Hessian approaches zero via a saddle node bifurcation, and cf. Fig.~\ref{lamE}. An example of the displacement field (which in this case is purely non-affine) is shown in Fig.~\ref{nonaffE} together with the eigenfunction associate with the lowest eigenvalue. The dot-product of the non-affine displacement and the eigenfunction $|\B u \cdot \B \Psi_1|$ in the present case is 0.998.
It can be observed that the non-affine event triggered by an increase in the electric field is not dipolar. While the quadrupolar structure is less pronounced than in the case of pure mechanical strain, it appears that the elastic interactions have the upper hand in the quasi-localized events also in the present case. 

\begin{figure}
	\includegraphics[width=0.70\columnwidth]{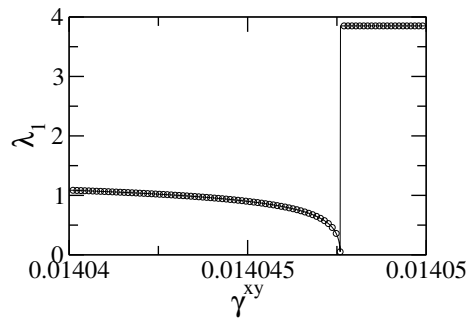}
	\caption{A typical change of the lowest non-zero eigenvalue associated with a plastic event due to increase in strain in the presence of an electric field. Here the field is $E^x=0.5$ and the steps of strain are $\Delta \gamma^{xy}=10^{-8}$.}
	\label{lamgame}
\end{figure}

\begin{figure}
	\includegraphics[width=0.75\columnwidth]{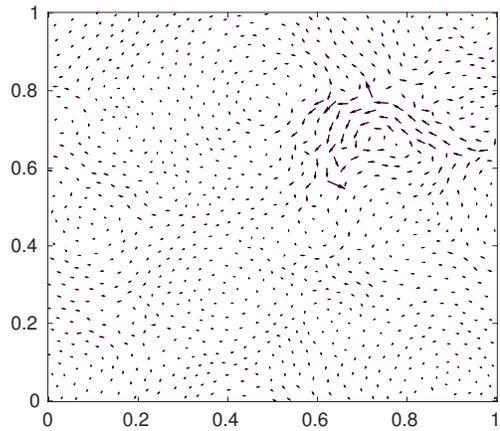}
	\caption{A typical non-affine response upon the increase of strain in the presence of an electric field. The non-affine response is shown together with the eigenfunction associated with the lowest non-zero eigenvalue, cf. Fig.~\ref{lamgame}. The dot-product of the non-affine displacement and the eigenfunction $|\B u \cdot \B \Psi_1|$ in the present case is 0.999. Again the quadrupolar nature of the event is less pronounced than in the case of pure mechanical strain, but it is still not a dipolar event.}
	\label{nonaffgame}
\end{figure}

Finally we examine the plastic events occurring when strain is increased in the presence of electric field. The lowest non-zero eigenvalue as a function of strain and the non-affine response just before the plastic event are shown in Figs.~\ref{lamgame} and \ref{nonaffgame} respectively. 

\section{Theory of moduli}
\label{moduli}

The moduli that are of interest in this system are all tensors, but for simplicity we will drop tensorial notation. We demonstrate the theory of the moduli using the two that determined the polarization as a function of strain and electric field, i.e. the piezoelectric coefficient $\Sigma$ and the polarizability $\chi$: 
\begin{equation}
 P^x = \Sigma \gamma^{xy}\ , \quad  P^x = \chi  E^x \ .
 \label{mod2}
\end{equation}
Of course other components and other moduli can be defined similarly, but the theory repeats along the lines shown below. 

As always, the moduli have an affine and a non-affine contribution \cite{11HKLP}. To write them explicitly we need to solve first for the non-affine displacement field. This field is obtained from the requirement of mechanical equilibrium, stating that the force on each particle $\B f_i \equiv \sum_j \B f_{ij} + q_i \B E$ needs to vanish before and after every increase in strain or electric field. We thus write
\begin{equation}
\frac{ d \B f_i}{d\B \gamma} =0\ , \quad \frac{d \B f_i}{d\B E} =0 \ .
\end{equation}
To proceed we remember that $\B f_i \equiv -\partial U/\partial \B r_i$. In addition, during straining $\B r_i$ has an affine change, for example Eq.~(\ref{simpleshear}) and a non affine step $\B u_i$. Then the condition for equilibrium reads
\begin{equation}
-\frac{d}{d\B \gamma} \frac{\partial U}{\partial \B r_i} = -\frac{d}{d \B E} \frac{\partial U}{\partial \B r_i} =0 \ .
\end{equation}
These conditions translate to the equations
\begin{eqnarray}
-\B \Xi_i -\sum_j \B H_{ij} \frac{d\B u_j}{d\B \gamma}&=& 0 \ ,\\	
\B q - \sum_j \B H_{ij} \frac {d \B u_j}{d\B E}&=&0 \ ,
\end{eqnarray} 
where $\B \Xi_i$ is the ``non-affine force" $ \frac{\partial^2 U}{\partial \B r_i\partial \B \gamma}$  and $\B q$ is a vector of charges of the particles. 

Consider then the polarizability $\chi$ of Eq.~(\ref{mod2}):
\begin{equation}
\chi \equiv \frac{1}{L^2}\frac{d \sum_i q_i r_i^x}{dE^x} = \frac{1}{L^2}\sum_i q_i \frac{d u_i^x}{dE^x} = \frac{1}{L^2}\sum_{i,j} q_i \B H_{ij}^{-1} \B q_j \ .
\label{chi}
\end{equation} 

Similarly we write for the piezoelectric coefficient  
\begin{eqnarray}
\Sigma &\equiv& \frac{1}{L^2}\frac{d \sum_i q_i r_i^x}{d\gamma^{xy}} = \frac{1}{L^2}\left[\sum_i q_i \frac{d u_i^x}{d\gamma^{xy}} + \sum_i q_i r_i^y\right]\nonumber \\
&=& -\frac{1}{L^2}\sum_{i,j} q_i \B H_{ij}^{-1} \B \Xi_j + P^y\ .
\label{Sigma}
\end{eqnarray} 
The reader should note that initially, without shear or electric field $P^x$ and $P^y$ should vanish in the thermodynamic limit. Since we simulate small systems, contributions like $P^y$ in the last equation are finite and in fact non negligible at all. Even though this is a finite size effect, comparing theory with simulations these contributions are important. 

In table \ref{table1} we validate the theory for the polarizability as shown in Eq.~(\ref{chi}). The slope of $P^x$ vs. $E^x$ is computed directly from the data, cf. Fig.~\ref{E} panel (c). The slope for different value of $E^x$ is compared with the theoretical prediction Eq.~(\ref{chi}).   
\begin{table}[h!]
	\begin{center}
		\caption{Polarizability from simulation and theory.}
		\label{table1}
		\begin{tabular}{l|c|r}
			$E^x$ & Slope of $P^x$ vs. $E^x$ & Theory\\
			\hline
			0.5 & $6.18738\times 10^{-4}$ & $6.18405\times 10^{-4}$ \\ \hline
			3.0 & $3.57812\times 10^{-4}$ & $3.57507\times 10^{-4}$ \\ \hline
			5.5 & $3.02051\times 10^{-4}$ & $3.02004\times 10^{-4}$ \\ \hline
			8.0 & $3.54587\times 10^{-4}$ & $3.54383\times 10^{-4}$ \\ \hline
			10.5 & $3.65511\times 10^{-4}$ & $3.65198\times 10^{-4}$ \\ \hline
			12.5 & $3.51213\times 10^{-4}$ & $3.50964\times 10^{-4}$ \\ \hline
			14.0 & $3.12568\times 10^{-4}$ & $3.12398\times 10^{-4}$ \\ \hline
			18.5 & $3.45697\times 10^{-4}$ & $3.45082\times 10^{-4}$ \\ \hline
			23.5 & $3.51897\times 10^{-4}$ & $3.82520\times 10^{-4}$ \\ \hline
			28.0 & $6.02561\times 10^{-4}$ & $5.97835\times 10^{-4}$ \\ \hline
		\end{tabular}
	\end{center}
\end{table}
We can see that the numbers agree up to errors of the order of $10^{-7}	$. In our view this is an excellent agreement between theory and simulations. One should note that computing slopes from data has always some numerical errors. In table \ref{table2} we compare the measured value of the piezoelectric coefficient $\Sigma$ from the simulations, cf Fig.~\ref{withE} panel (c), with the theory of Eq.~(\ref{Sigma}). Indeed, we learn that the numerical value of $P^y$ in our finite system is of the same order as the contribution of the non-affine term. Taken together as requested by Eq.~(\ref{Sigma}) we find again excellent agreement with the numerical slope up to errors of the order of $10^{-4}-10^{-3}$. 

One can define and measure other moduli, but the agreement found here indicates that we have the appropriate theory under hand, and any desired modulus can be computed as shown here using the same techniques. We did check that the Maxwell relation between the piezoelectric coefficient and the electrostriction coefficient (i.e. the slope of stress due to increase in electric field) is satisfied to very high accuracy. 
\begin{widetext}
\begin{center}
\begin{table}[h!]
		\caption{Piezoelectric coefficient from simulation and theory.}
		\label{table2}
		\begin{tabular}{l|c|c|c|r}
			$\gamma^{xy}$ & Slope of $P^x$ vs. $\gamma^{xy}$ & $P^y$(A) & non-affine part(B) &A+B \\ \hline
			0.005 & $-1.19916\times 10^{-1}$ & $-4.60709\times 10^{-2}$ & $-5.35668\times 10^{-2}$ & $-9.96377\times 10^{-2}$ \\ \hline
			0.016 & $-4.78695\times 10^{-2}$ & $-4.64870\times 10^{-2}$ & $-1.42070\times 10^{-3}$ & $-4.79077\times 10^{-2}$ \\ \hline
			0.027 & $-6.96405\times 10^{-2}$ & $-4.77154\times 10^{-2}$ & $-2.19214\times 10^{-2}$ & $-6.96368\times 10^{-2}$ \\ \hline
			0.035 & $-5.41307\times 10^{-2}$ & $-4.76207\times 10^{-2}$ & $-5.84410\times 10^{-3}$ & $-5.34648\times 10^{-2}$ \\ \hline
			0.042 & $-3.69254\times 10^{-3}$ & $-4.64967\times 10^{-2}$ & $4.300644\times 10^{-2}$ & $-3.49026\times 10^{-3}$ \\ \hline
			0.048 & $-9.40170\times 10^{-3}$ & $-4.64093\times 10^{-2}$ & $3.682565\times 10^{-2}$ & $-9.58365\times 10^{-3}$ \\ \hline
			0.056 & $-2.64813\times 10^{-2}$ & $-4.72040\times 10^{-2}$ & $2.118630\times 10^{-2}$ & $-2.60177\times 10^{-2}$ \\ \hline
			0.064 & $-5.75430\times 10^{-2}$ & $-4.73491\times 10^{-2}$ & $-1.04216\times 10^{-2}$ & $-5.77707\times 10^{-2}$ \\ \hline
			0.074 & $-1.08118\times 10^{-1}$ & $-4.62390\times 10^{-2}$ & $-6.22780\times 10^{-2}$ & $-1.08517\times 10^{-1}$ \\ \hline
			0.088 & $-1.57305\times 10^{-2}$ & $-4.49603\times 10^{-2}$ & $2.905880\times 10^{-2}$ & $-1.59015\times 10^{-2}$ \\ \hline
			0.094 & $-3.45504\times 10^{-2}$ & $-4.49485\times 10^{-2}$ & $1.188220\times 10^{-2}$ & $-3.30663\times 10^{-2}$ \\ \hline
		\end{tabular}
\end{table}
\end{center}
\end{widetext}

\section{summary and conclusions}
\label{summary} 

In summary, we have analyzed in some detail the responses of an assembly of charged particles to mechanical and electric external strains. Even though the electric interactions are long ranged, we learn that all the non-affine responses are dominated by the short range elastic forces, displaying quasi-localized non-affine displacement fields that do not reflect the existence of electric long-ranged forces. One could naively think that increasing for example the charges in each grain, the electrostatic interaction will become more dominant. In fact increasing the charges only makes the system more compressed, and in balance the short range forces increase and dominate again.  Even though the Hessian matrix does not explicitly depend on the electric field, the fact that the forces do, make the presence of electric field very important in determining the non-affine displacement fields. The Hessian formalism provides a transparent and precise theory for computing the interesting moduli of electrostriction and piezoelectric coefficients. All the plastic events that were observed stemmed from a saddle node bifurcation with an eigenvalue that goes to zero with a square-root singularity.

It is important to note that we have considered a very specific model with disks that cannot deform, whose charge is fixed to their center of mass. In general, especially if consider real dielectric grains, then the responses to external strains and electric field will include charge redistribution and deformation. In future work it would be relevant and extend the present work to experimental systems like charged colloids and granular assemblies with surface charges for which these effects will be important. 

\bibliography{ALL}

\end{document}